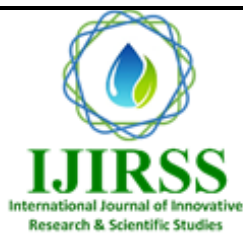
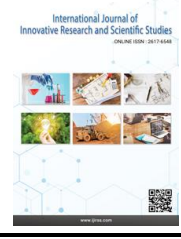

# The future of gravitational wave science – unlocking LIGO's potential: AI-driven data analysis and exploration

Yong Xiao, Li[1], Zin Nandar Win[1*], He Wang[2], Hla Myo Tun[3], Win Thu Zar[3]

[1]School of Electronic Engineering, Beijing University of Posts and Telecommunications, Beijing 100867, China.
[2]International Centre for Theoretical Physics Asia-Pacific (ICTP-AP), University of Chinese Academy of Sciences (UCAS), Beijing 100049, China.
[2]Taiji Laboratory for Gravitational Wave Universe (Beijing/Hangzhou), University of Chinese Academy of Sciences (UCAS), Beijing 100049, China.
[3]Department of Electronic Engineering, Research Department and International Relations Office, Yangon Technological University, Yangon 11011, Myanmar.

Corresponding author: Zin Nandar Win (*Email: zin.nandar.win@bupt.edu.cn*)

## Abstract

The advent of gravitational wave astronomy (GW) has revolutionized the observation of cataclysmic cosmic events, such as black hole mergers and neutron star collisions. The Laser Interferometer Gravitational-Wave Observatory (LIGO) has been at the forefront of these discoveries. However, the immense volume and complexity of gravitational wave data present significant challenges for traditional analysis methods. This paper investigates the growing synergy between artificial intelligence (AI) and GW science, emphasizing how AI enhances signal detection, noise reduction, and data interpretation. It begins with an overview of GW fundamentals and the role of machine learning in increasing detector sensitivity. Notable GW events observed by LIGO are discussed alongside persistent analytical challenges such as data quality, generalization, and computational constraints. A comprehensive performance review of AI techniques, including supervised learning, unsupervised learning, deep learning, and reinforcement learning is presented based on data spanning 2021 to 2024. Evaluation metrics include accuracy, precision, true positive rate (TPR), false positive rate (FPR), and computational efficiency. Findings indicate that deep learning and supervised learning outperform other approaches, particularly in enhancing TPR and minimizing FPR. While unsupervised and reinforcement learning models offer less precision, they demonstrate high efficiency and potential for real-time applications. The study also explores AI's integration into next-generation detectors and waveform reconstruction techniques. Overall, the integration of AI into GW research significantly improves the reliability and speed of event detection, unlocking new possibilities for exploring the dynamic universe. This paper provides a comprehensive outlook on the transformative role of AI in shaping the future of GW astronomy.

**DOI:** 10.53894/ijirss.v8i3.7514
**Funding:** This research was supported by [National Science Foundation for Young Scholars of China] under (Grant Number: 62205031). Additional support was provided by Beijing University of Posts and Telecommunications.
**History:** Received: 2 April 2025 / **Revised:** 7 May 2025 / **Accepted:** 9 May 2025 / **Published:** 30 May 2025

**Competing Interests:** The authors declare that they have no competing interests.
**Authors' Contributions:** All authors contributed equally to the conception and design of the study. All authors have read and agreed to the published version of the manuscript.
**Transparency:** The authors confirm that the manuscript is an honest, accurate, and transparent account of the study; that no vital features of the study have been omitted; and that any discrepancies from the study as planned have been explained. This study followed all ethical practices during writing.
**Acknowledgements:** The authors express sincere gratitude to Beijing University of Post and Telecommunications for providing research facilities and a conducive environment and also deeply thankful to the LIGO Scientific Collaboration and Virgo Collaboration for providing access to gravitational wave data and resources, which were essential for this study. Their groundbreaking work in gravitational wave detection has been a constant source of inspiration.
**Publisher:** Innovative Research Publishing

## 1. Introduction

Gravitational waves (GWs), ripples in the fabric of space-time first predicted by Einstein [1], were directly detected for the first time in the LIGO Scientific Collaboration and Virgo Collaboration [2] by the Laser Interferometer Gravitational-Wave Observatory (LIGO). This landmark discovery, involving the merger of two black holes (GW150914), confirmed Einstein's century-old theory and opened a whole new observational window on the universe. Since then, the detection of various gravitational wave events, including binary black hole mergers, neutron star collisions, and multi-messenger phenomena such as GW170817, has profoundly changed our understanding of the cosmos [3, 4]. These discoveries, made possible by advanced detectors such as LIGO, Virgo, and KAGRA, have greatly expanded the scope of observational astronomy, allowing scientists to observe cataclysmic cosmic events in unprecedented detail.

LIGO, Virgo, and KAGRA key observatories in the study of gravitational waves are responsible for detecting a range of astrophysical phenomena once thought to be undetectable. LIGO's first observing run (O1) in 2015 marked a pivotal moment with the detection of GW150914, the first confirmed binary black hole merger. Subsequent observing runs, including O2 and O3, led to groundbreaking discoveries such as the binary neutron star merger GW170817, which provided the first multi-messenger astrophysical signal and confirmed the production of heavy elements such as gold and platinum [4]. Those detections were followed by other discoveries in the O3 run, including GW 190412, the first detection of a binary black hole merger with unevenly massive components, and the discovery of intermediate-mass black holes, including GW 190521.

However, as the sensitivity of gravitational wave detectors continues to improve, the volume and complexity of the data generated have overwhelmed traditional analysis methods, posing significant challenges. This growing data influx necessitates the use of artificial intelligence (AI) and machine learning (ML) technologies, which offer novel approaches to processing and analyzing massive data. AI-driven approaches, particularly deep learning algorithms such as convolutional neural networks (CNNs) and recurrent neural networks (RNNs), have proven effective in overcoming challenges such as noise reduction, real-time signal detection, and waveform reconstruction [5, 6]. These techniques enable the identification of faint signals buried in noise, improving detector sensitivity and accelerating the discovery of new astrophysical phenomena.

Despite this progress, there are still significant challenges to making AI models accurate and scalable, especially related to data quality, computational efficiencies, and model generalization [7]. As next-generation detectors, such as the Einstein Telescope and Cosmic Explorer, will generate even larger amounts of data, AI will play a critical role in managing these data streams and addressing the complexities associated with future gravitational wave detection.

The aim of this paper is to provide a comprehensive exploration of the intersection between artificial intelligence (AI) and gravitational wave science, focusing on how AI methods are being integrated into gravitational wave analysis. It reviews current trends, technological developments, and challenges, and considers ways in which AI can optimize the use of sophisticated detectors and facilitate the analysis of increasing amounts of data. It also discusses the contributions of LIGO, Virgo, and KAGRA to the field and looks ahead to future opportunities offered by next-generation observatories, underscoring and highlighting the transformative potential of AI in shaping the future of gravitational wave astronomy.

## 2. Gravitational Wave's Notable Discovered Events Discovered in LIGO Detectors

*2.1. LIGO 1st Observing Run O1 (September 2015-January 2016)*

Advanced LIGO's first observing run (O1), which lasted four months, represented a significant step forward in astrophysics. This was demonstrated by the identification of three significant gravitational wave (GW) events, each of which provided a unique addition to the emerging field of GW astronomy. The initial event, GW150914, holds particular importance as it marked the first direct detection of gravitational waves. This offered conclusive evidence of binary black hole mergers and marked the beginning of a new era in observational astronomy. This ground breaking discovery was succeeded by LVT15012, a candidate event that further corroborated the detection capabilities of Advanced LIGO and indicated that such cosmic occurrences could be observed with greater regularity. The third incident, GW151226, marked the second confirmed discovery of gravitational waves, distinguished by lower-mass black holes in contrast to GW150914. This observation was highly significant, as it illustrated the potential for identifying gravitational waves across various black hole masses. This observation further solidified the conventional view that black hole mergers are significant generators of gravitational waves. The detections during O1 validated Einstein's general theory of relativity and established Advanced LIGO as a formidable instrument for investigating the universe via gravitational waves, thereby creating new opportunities for comprehending the dynamics of black holes and other astrophysical phenomena.

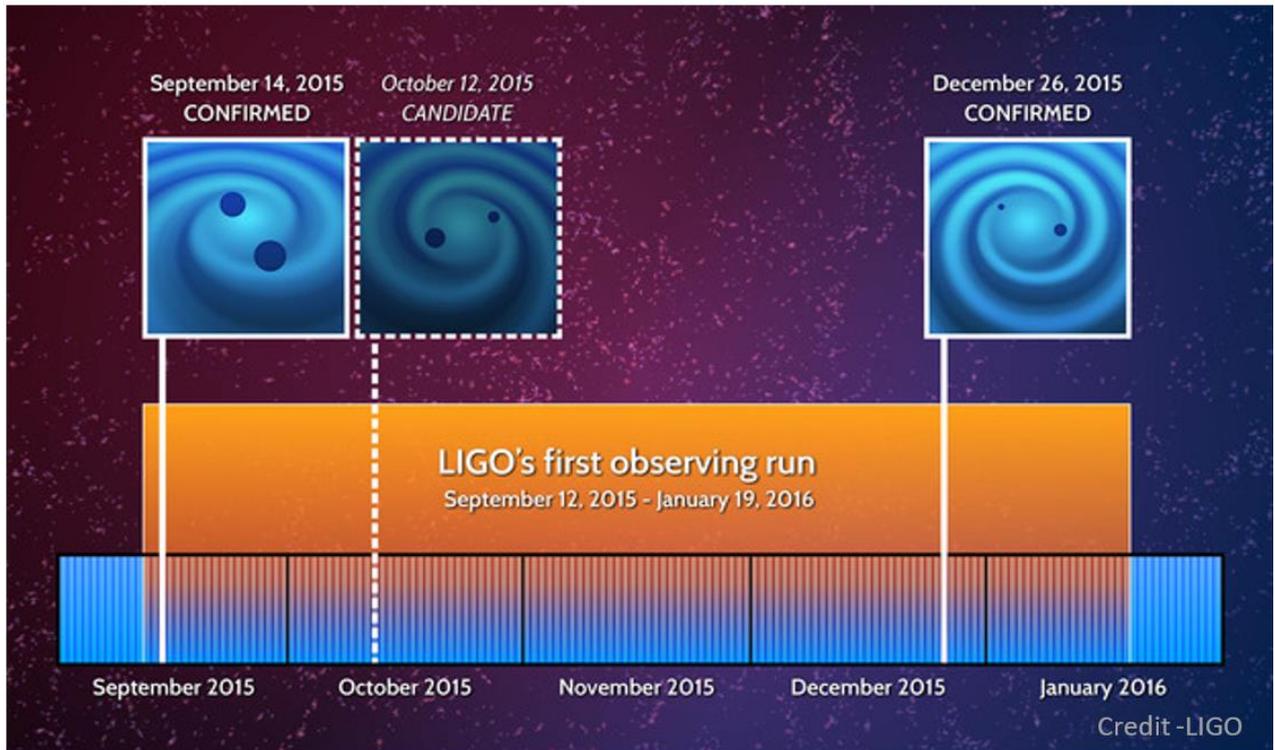

**Figure 1.**
Timeline of LIGO's first observing run (September 12, 2015 – January 19, 2016), showing the dates of two confirmed gravitational-wave detections and one candidate event.

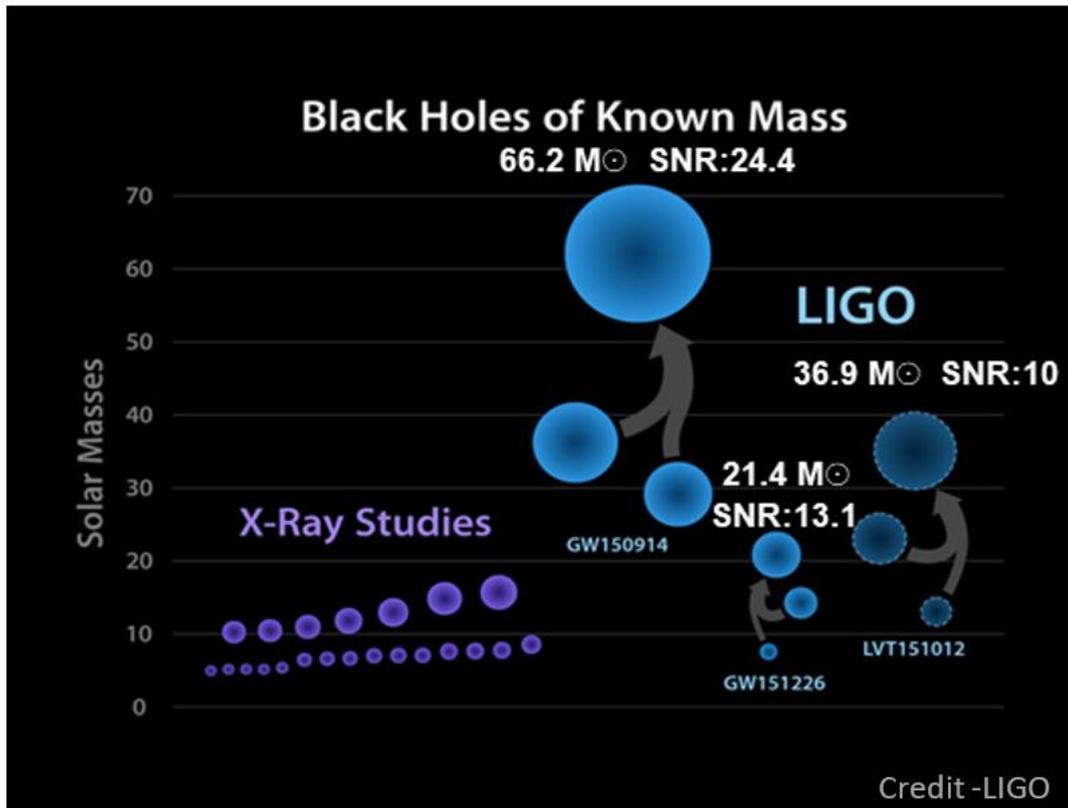

**Figure 1.**
Known black hole masses from X-ray studies and LIGO detections, highlighting LIGO's discovery of more massive black holes.

*2.2. LIGO 2$^{nd}$ Observing Run O2((November 2016- August 2017)*

During the second observing run of Advanced LIGO (Advanced Laser Interferometer Gravitational-Wave Observatory, O2, November 2016 - August 2017), several innovative gravitational-wave occurrences were discovered. These included GW170608, which revealed mergers with smaller-mass black holes, and GW170104, which was the first discovery of 2017 and confirmed more binary black hole mergers. The first binary neutron star merger observation, GW170817, with multi-messenger counterparts confirming the astrophysical source of heavy elements, and GW170814, the first event to be concurrently seen by LIGO and Virgo, thus enabling better source localization.

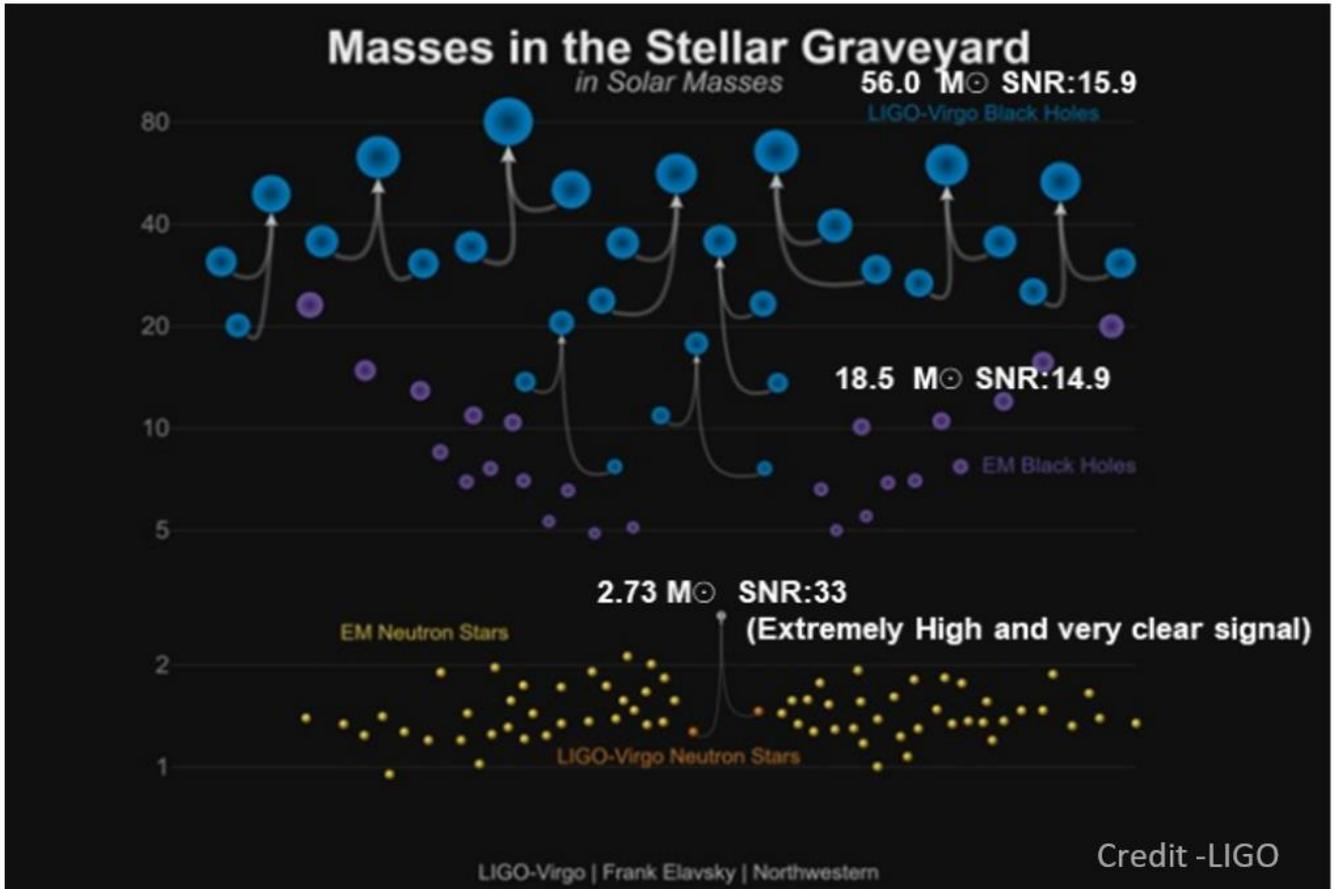

**Figure 2.**
Masses of black holes and neutron stars detected by LIGO/Virgo (blue, orange) and electromagnetic observations (purple, yellow), illustrating the diversity and mass range of compact objects in the universe.

*2.3. LIGO 3rd Observing Run O3: (Apr 2019 - Mar 2020)*

During the O3 observing run, several significant gravitational-wave events were reported, further enriching the landscape of compact object mergers. The observation of GW190412 represented the inaugural instance of a binary black hole merger with components of unequal masses, thereby offering novel insights into the processes of black hole formation. GW190425 was the second detection of a binary neutron star merger, and it is notable for its substantially higher total mass compared to GW170817. GW190521 was the first intermediate-mass black hole (IMBH) merger to be observed, with a large total mass and a short, low-SNR signal due to its distance and mass, posing additional challenges for detection. GW190814 exhibited a distinctive merger between a black hole and a second object whose mass lies within the so-called "mass gap," thereby rendering its classification (as either the heaviest neutron star or the lightest black hole) ambiguous, with the event distinguished by its high signal-to-noise ratio. Finally, GW200105 and GW200115 provided the first and second confirmed detections of neutron star–black hole mergers (NSBH), further confirming the existence of such systems, though no associated electromagnetic counterpart was observed in either case.

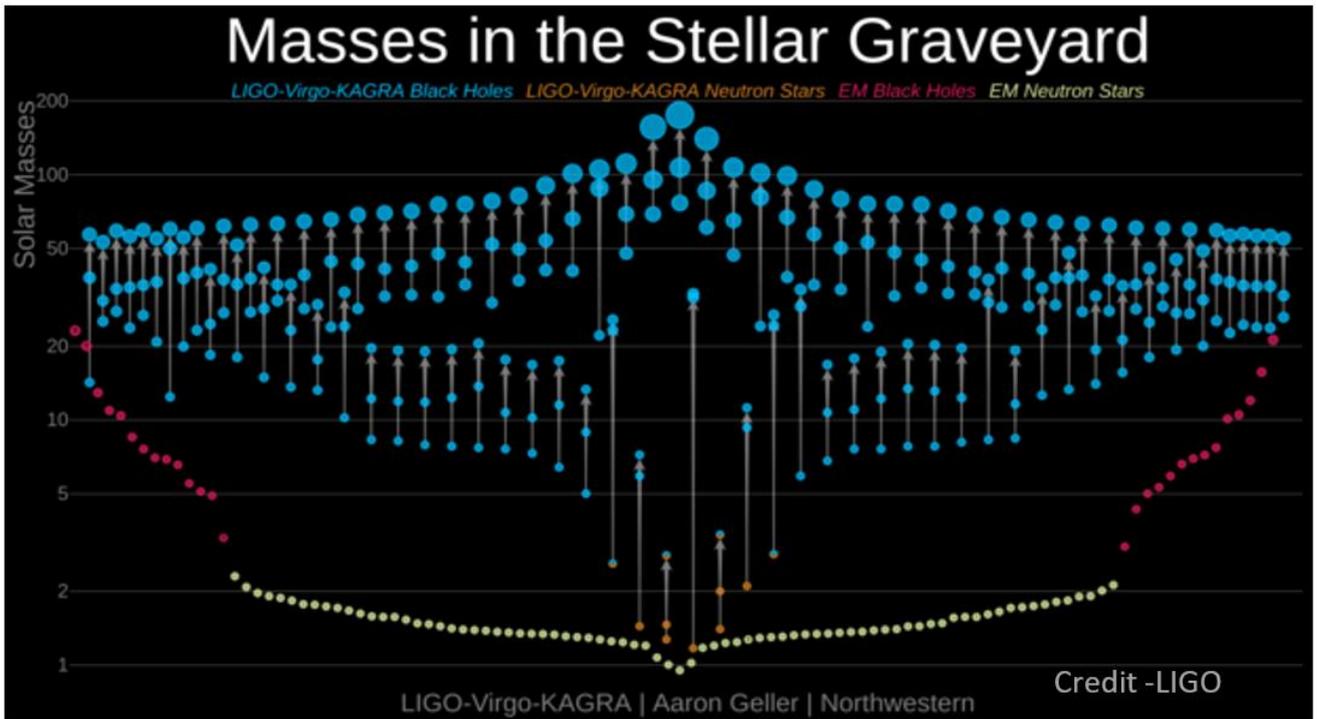

**Figure 3.**
Graphic of masses of announced gravitational-wave detections and black holes and neutron stars previously constrained through electromagnetic observations. This version shows all events through the end of O3 with a probability of astrophysical origin p_astro > 0.5.

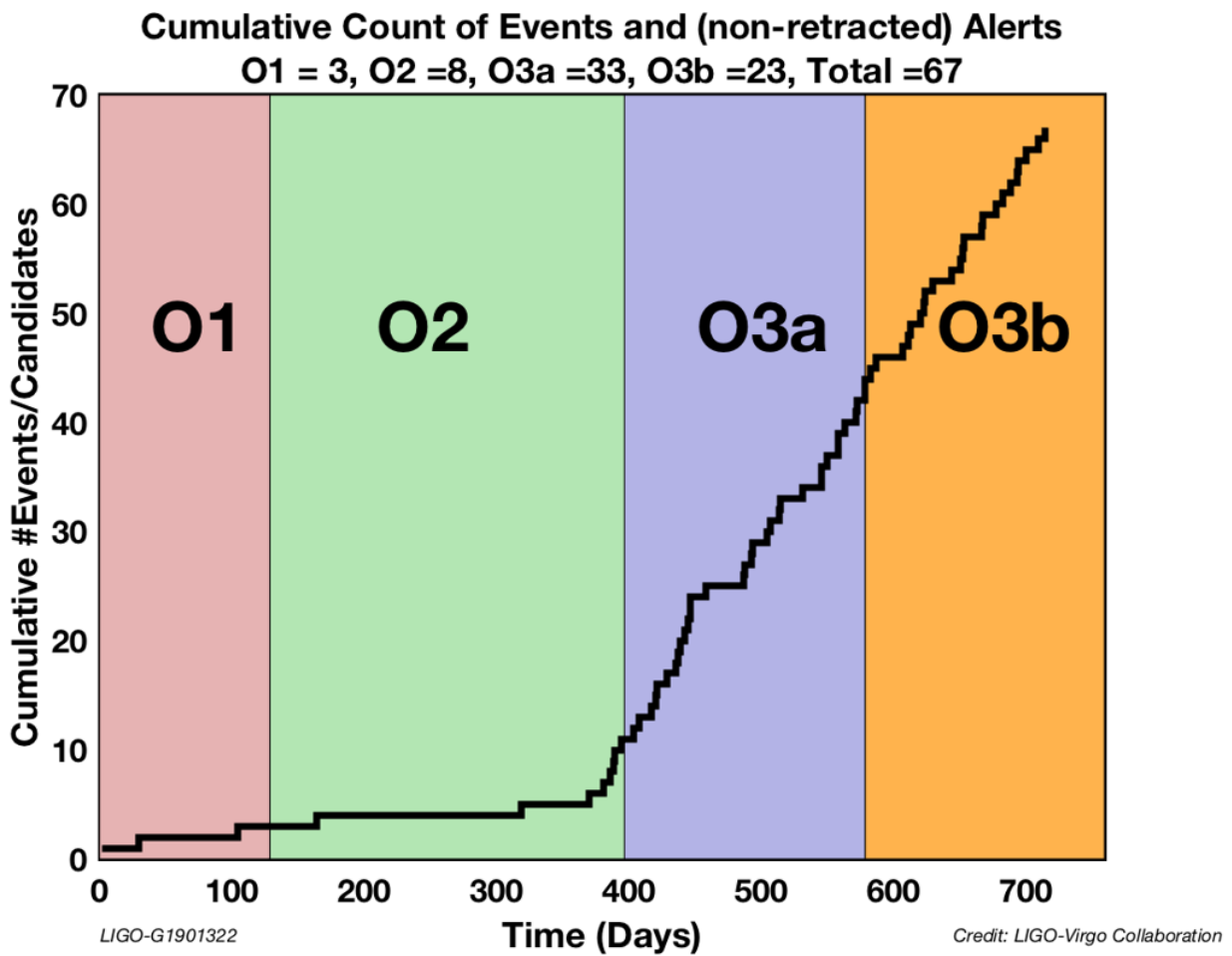

**Figure 4.**
LIGO's third observing run yielded 56 gravitational-wave detections. That's 5-times more than were collectively made during the first two observing runs of the Advanced detector era (O1 and O2 yielded 11 gravitational wave detections). Image credit: LIGO.

*2.4. LIGO 4th Observing Run O4: (May 2023~Current)*

The fourth observing run of the Laser Interferometer Gravitational-Wave Observatory (LIGO) was extended in February 2025. Now, its end is expected in October 2025. Regarding the time lost from the anticipated April/May 2025 hiatus and unplanned interruptions in 2024, it is obvious that this delay will generate a comparable amount of calendar observation time to that envisioned with the prior June 2025 finish date.

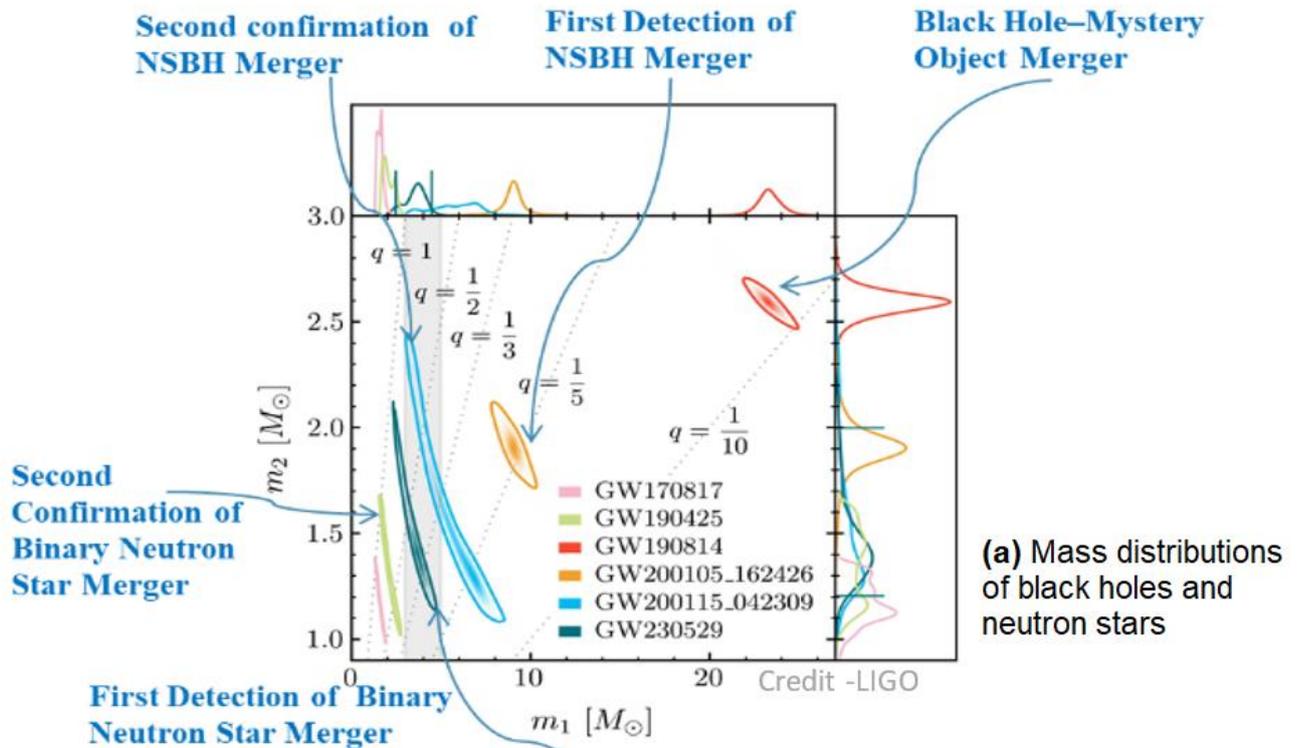

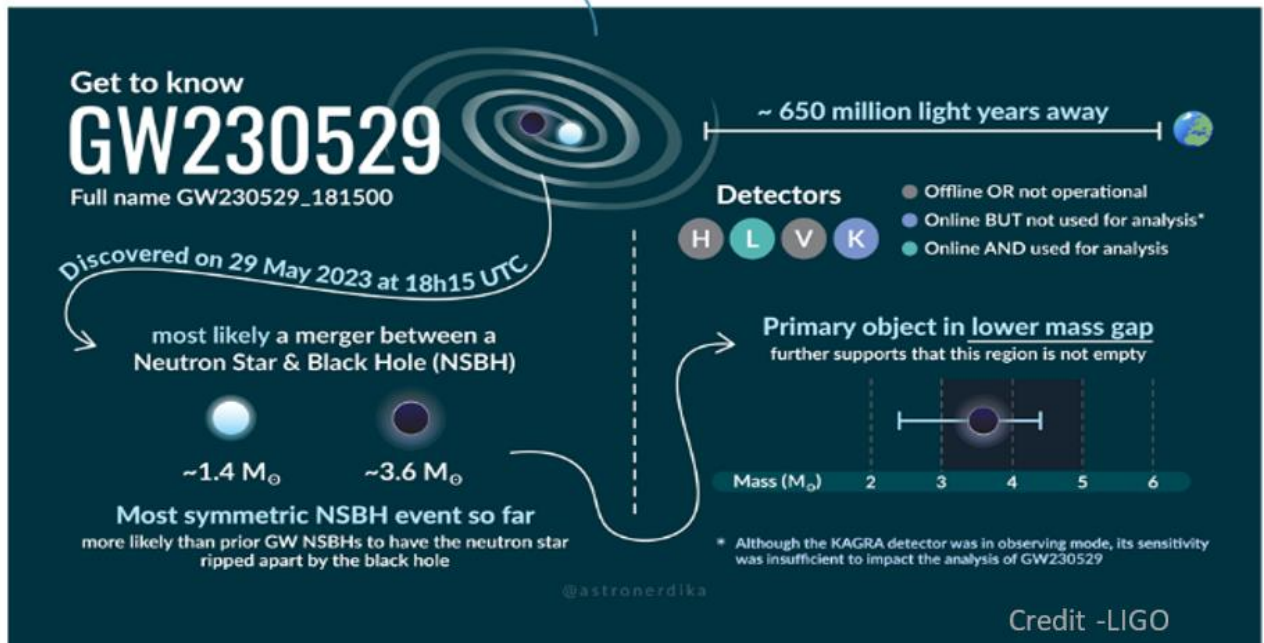

(b) Summary of key neutron star–black hole mergers

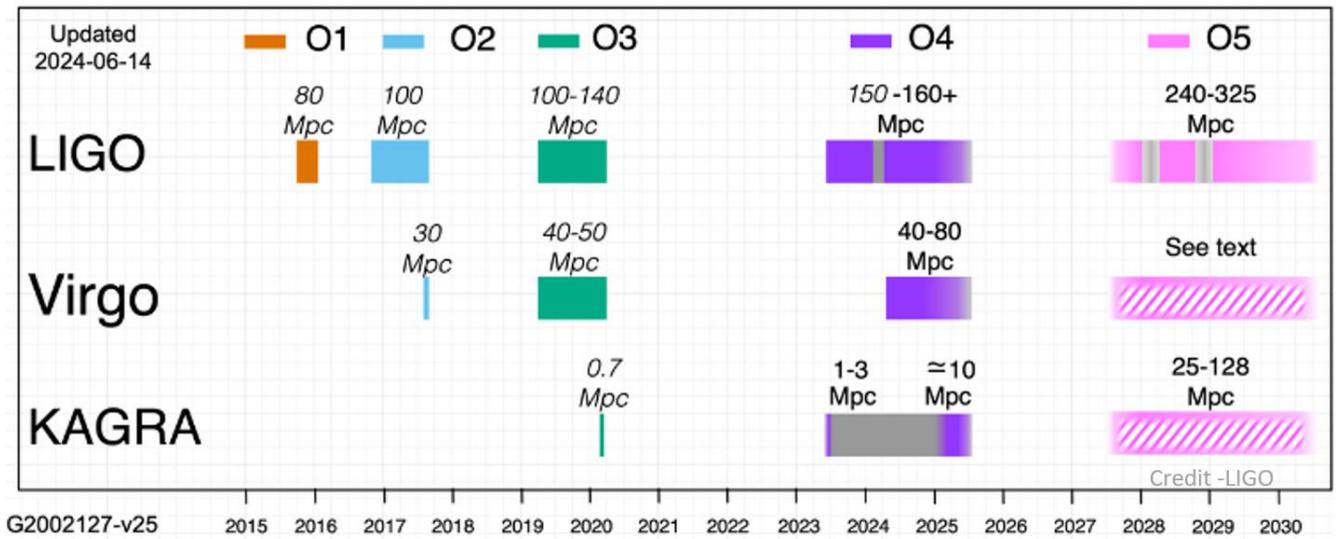

(c) Current timeline for the Observing run.

**Figure 5.**
LIGO 4th Observing Run O4(May'23~Current): (a)Mass distributions of black holes and neutron stars from notable LIGO/Virgo gravitational wave events.(b) Summary of key neutron star–black hole mergers, highlighting GW230529 as the most symmetric event with a primary object in the lower mass gap.(c) Current (as of June 2024) timeline for the O4 observing run.

*2.5. LIGO Laboratory Statement on Long Term Current and Future Observing Plans*

Since 2015, the U.S. National Science Foundation Laser Interferometer Gravitational-Wave Observatory (NSF LIGO) Hanford and Livingston Observatories have conducted numerous successful observations in cooperation with the LIGO Scientific Collaboration, the European Gravitational Observatory, and the Virgo Collaboration, employing the Advanced NSF LIGO interferometers. Run 4, O4, was first detected by the Hanford and Livingston interferometers on May 24, 2023. A series of enhancements follows O4. LIGO, Virgo, and KAGRA will take part in a long observational campaign, O5, expected to run until roughly 2028 once these improvements have been completed.

The LIGO Laboratory is completely committed to conducting continuous research on the gravitational-wave universe after 2028. Apart from O5, LIGO is now planning interferometer enhancements as well as observational phases into the decade that follows. These strategies will keep evolving in response to comments from the LIGO-Virgo-KAGRA gravitational-wave community and external conditions, perhaps undergoing further development and refinement. They are also dependent on LIGO Operations funds being accessible.

## 3. Materials and Methods
*3.1. Gravitational Waves in General Relativity*

General relativity predicts the existence and propagation of gravitational waves as solutions to the Einstein field equations. We present a detailed derivation of the wave equation and discuss the tensor nature of gravitational waves. The two fundamental polarizations, known as the plus and cross polarizations, are introduced, and their mathematical representations are derived. The relationship between the polarization and the source's quadrupole moment is established, highlighting how changes in the source's mass distribution lead to the emission of gravitational waves with specific polarizations.

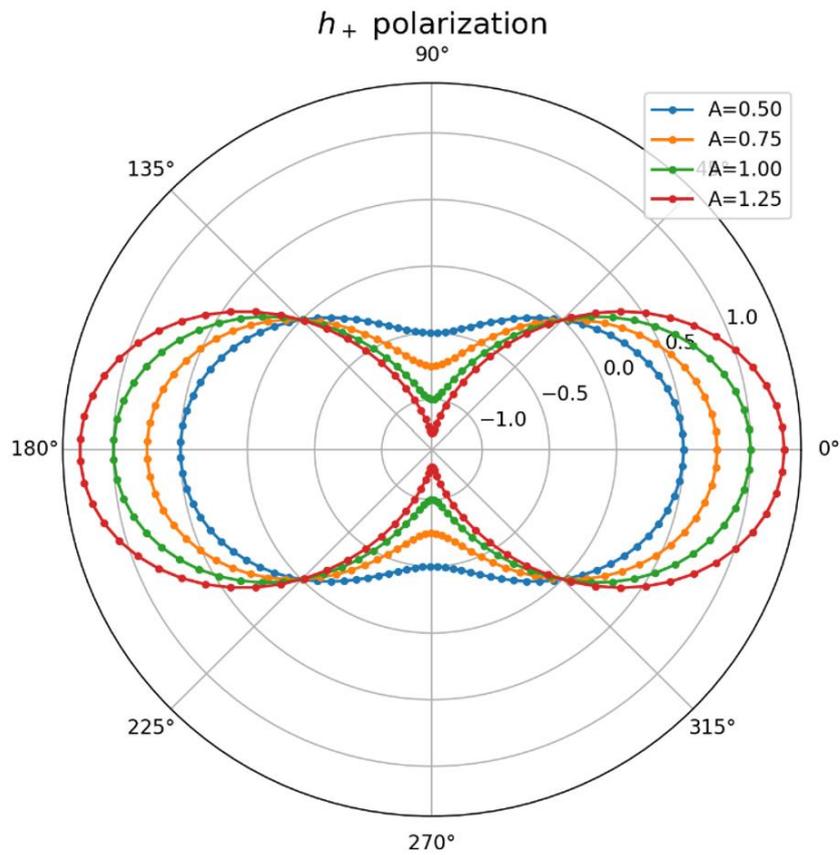

**Figure 7.**
h₊ polarization of Gravitational Waves.

*3.2. The Spin-2 Graviton in Quantum Field Theory*

The spin-2 nature of the graviton is shown to be consistent with the symmetry properties of the gravitational field. The Feynman rules for graviton interactions are formulated, and the role of the graviton in mediating gravitational forces between particles is explained. The challenges in quantizing the gravitational field and the resulting theoretical ambiguities are also discussed.

The close link between the spin-2 character of the graviton and the polarizing effect of gravitational waves. The tensor polarization of the gravitational waves is shown to be a direct expression of the spin-2 nature of the graviton. A unified hypothetical viewpoint is offered by explaining the principles of conservation rules and symmetry principles underlying this link.

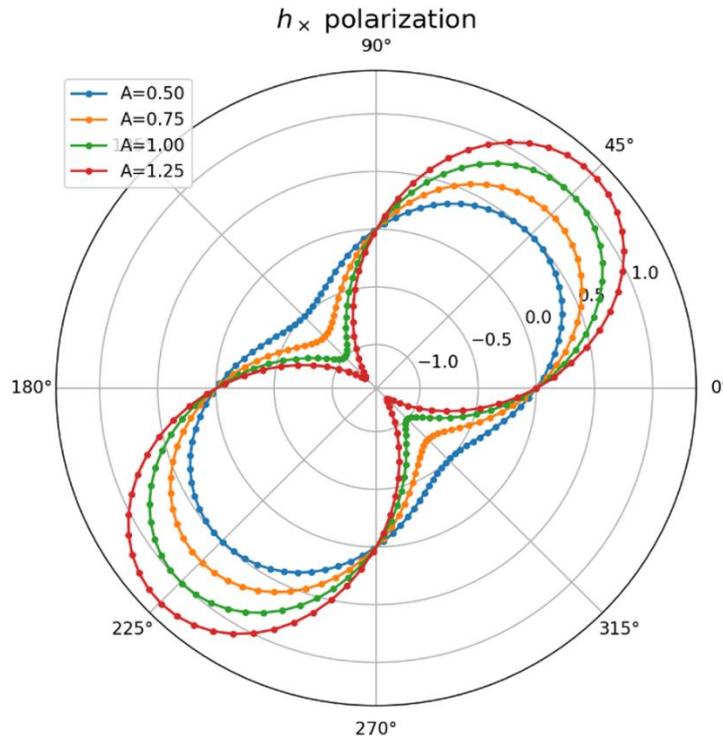

**Figure 8.**
$h_x$ polarization of Gravitational Waves.

Our theoretical investigations and simulations of gravitational wave sources have yielded significant results regarding the polarity of gravitational waves. In the case of binary black hole mergers, we have found that the plus and cross polarizations exhibit distinct patterns as predicted by general relativity. The amplitude and phase evolution of these polarizations during the inspiral, merger, and ringdown phases have been precisely calculated.

$$h_{\mu\nu} = \frac{16\pi G}{c^4} \tau_{\mu\nu} \qquad [1]$$

where $h_{\mu\nu}$ is the wave operator and $G$ is Newton's gravitational constant. The quantity $\tau_{\mu\nu}$ plays the role of an effective energy-momentum tensor, containing the actual energy-momentum tensor.

$$\tau_{\mu\nu} = |g| T_{\mu\nu} + \frac{c^4}{16\pi G} \Lambda_{\mu\nu} \qquad [2]$$

Where $|g|$ is the determinant of the metric. The quantity $\tau_{\mu\nu}$ plays the role of an effective energy-momentum tensor, containing the actual energy-momentum tensor $T_{\mu\nu}$, together with a quantity $\Lambda_{\mu\nu}$ that contains all the trems non-linear in $h_{\mu\nu}$.

For example, during the inspiral phase, the amplitude of the plus polarization grows steadily as the two black holes approach each other, while the cross polarization also shows a correlated but phase-shifted behavior. The merger event leads to a sharp peak in both polarizations, followed by a characteristic dampening oscillation during the ringdown phase.

Moreover, our studies on neutron star mergers have revealed additional complexity in polarization patterns. The presence of matter and the strong magnetic fields associated with neutron stars introduce modulations in the gravitational wave polarizations. We have identified specific frequency-dependent signatures in the polarizations that could potentially be used to distinguish neutron star mergers from binary black hole mergers and to gain insights into the internal structure and composition of neutron stars.

*3.3. Data Sources*

The performance metrics for various AI approaches in gravitational wave analysis were collected from a comprehensive study spanning the years 2021 to 2024. The data includes metrics such as accuracy, precision, false positive rate (FPR), true positive rate (TPR), and computational efficiency for four AI approaches:
1. Supervised Learning.
2. Unsupervised Learning.
3. Deep Learning.
4. Reinforcement Learning.

*3.4. Performance Metrics*

The following metrics were used to evaluate the performance of each AI approach:

Accuracy: The proportion of correctly classified instances.

Precision: The proportion of true positive predictions among all positive predictions.

FPR (False Positive Rate): The proportion of false positive predictions among all actual negatives.

TPR (True Positive Rate): The proportion of true positive predictions among all actual positives.

Computational Efficiency: A qualitative measure of the computational resources required, categorized as Low, Moderate, or High.

*3.5. Performance Comparison*

Rank AI Approaches: Rank the AI approaches based on their performance for each metric (e.g., Accuracy, Precision, TPR, FPR).

Identify Best Approach: Determine which AI approach performs best overall or for specific metrics.

Computational Efficiency: Compare the computational efficiency (Low, Moderate, High) of each AI approach and discuss trade-offs between performance and efficiency.

Based on the analysis, recommend the most suitable AI approach for gravitational wave analysis, considering both performance and computational efficiency.

**Table 1.**
Abbreviations and Definitions in AI-Powered Gravitational Wave Science.

| Abbreviation | Definition |
|---|---|
| GW | Gravitational Wave |
| LIGO | Laser Interferometer Gravitational-Wave Observatory |
| Virgo | A gravitational wave detector located in Italy, part of the global network for detecting gravitational waves. |
| KAGRA | A Japanese gravitational wave detector located in Japan, also part of the global network. |
| AI | Artificial intelligence |
| ML | Machine Learning |
| GW150914 | The first detection of gravitational waves, observed by LIGO, originated from the merger of two black holes on September 14, 2015. |
| GW151226 | A gravitational wave event detected by LIGO, originating from the merger of two black holes on December 26, 2015. |
| GW170104 | A gravitational wave event observed by LIGO on January 4, 2017, originating from the merger of two black holes. |
| GW170608 | A gravitational wave event detected by LIGO on June 8, 2017, from a binary black hole merger. |
| GW170814 | A gravitational wave event detected by LIGO and Virgo on August 14, 2017, from a binary black hole merger. |
| GW170817 | The first detection of a binary neutron star merger on August 17, 2017, observed by LIGO and Virgo. |
| GW 190412 | A gravitational wave event detected in 2019 from a binary black hole merger with unequal masses. |
| GW190425 | A binary neutron star merger event observed by LIGO and Virgo in April 2019. |
| GW 190521 | The first detection of an intermediate-mass black hole merger in May 2019. |
| GW200105 | A gravitational wave event from a binary black hole merger detected in January 2020. |
| GW200115 | A binary black hole merger detected by LIGO in January 2020. |
| LVT15012 | A candidate gravitational wave event detected by LIGO in 2015, later ruled out. |
| O1 | LIGO's first observing run (O1) in 2015 |
| O2 | The second observing run by LIGO and Virgo, from November 2016 to August 2017. |
| O3a | The third observing run of LIGO and Virgo, starting in April 2019 and continuing into 2020. |
| O3b | The second part of LIGO's third observing run (O3) |
| O4 | The fourth observing run for LIGO and Virgo which will further enhance detection sensitivity. |
| O5 | The fifth observing run for LIGO and Virgo, anticipated to start in the mid-2020s. |
| $M_\odot$ | Solar mass(1 $M_\odot$ = mass of the Sun). |
| Mpc | Megaparsec, a unit of distance used in astronomy (1 Mpc = 3.26 million light-years). |
| IMBH | Intermediate-mass Black Hole |
| SNR | Signal-to-Noise Ratio |
| NSBH | Neutron Star-Black Hole |
| BBH | Binary Black Hole |
| CNN | Convolutional Neural Networks |
| RNN | Recurrent Neural Networks |
| SVMs | Support Vector Machines |
| TPR | True Positive Rate |
| FPR | False Positive Rate |

## 4. Results and Discussions

The field of gravitational wave (GW) astronomy faces a multitude of challenges, particularly in leveraging artificial intelligence (AI) to enhance detection and analysis capabilities. One of the most pressing issues is data quality, severely impacted by pervasive noise contamination in GW detectors like LIGO and Virgo. These detectors are highly sensitive to environmental noise, such as seismic activity and wind, as well as instrumental noise like thermal fluctuations and laser instabilities. Traditional noise-filtering techniques often fall short due to the non-stationary, non-Gaussian nature of the noise, making it difficult to distinguish real GW signals from artifacts. Additionally, the variability in noise characteristics across different observatories complicates data calibration, requiring sophisticated methods to harmonize data from multiple detectors. The low signal-to-noise ratio (SNR) of weak GW signals further exacerbates the problem, as these signals are often buried under high levels of noise, making their detection and characterization particularly challenging. There is also an inherent trade-off between enhancing detector sensitivity and effectively rejecting noise, as increased sensitivity often amplifies noise sources, complicating the detection process. These data quality issues underscore the need for advanced AI-driven techniques, such as adaptive noise filtering and calibration methods, to improve the reliability and accuracy of GW detection. Another significant challenge lies in the computational demands of processing the vast amounts of data generated by advanced GW detectors. The sheer volume of data requires real-time processing and analysis to identify transient events like black hole mergers, as delays in processing can hinder rapid follow-up observations crucial for multi-messenger astronomy. Deep neural networks, while highly effective for tasks like signal detection and noise filtering, demand substantial computational resources due to their large number of parameters and complex architectures, raising concerns about scalability and efficiency. Processing latency further complicates matters, as the time required to analyze data can limit the ability to respond to astrophysical events in real time. Moreover, the complexity of waveform models used for tasks like matched filtering and parameter estimation imposes a high computational burden, particularly when analyzing signals from exotic sources or conducting large-scale searches across parameter spaces. These computational challenges highlight the need for optimized algorithms, distributed computing frameworks, and hardware acceleration to meet the growing demands of gravitational wave astronomy while maintaining the precision and speed required for groundbreaking discoveries. A third critical challenge is detection generalization, which refers to the ability of GW detectors to identify and characterize a wide range of astrophysical sources, each with distinct signal characteristics. These sources include binary black hole mergers, neutron star collisions, and potentially exotic phenomena like cosmic strings or primordial black holes. Variability in detection efficiency across detector networks and geographic locations further complicates this task, as differences in sensitivity, noise profiles, and operational conditions can affect the consistency of detections. Accurately localizing sources is another challenge, requiring precise triangulation of signals from multiple detectors, which is often hindered by uncertainties in signal arrival times and amplitudes. Traditional template matching methods, while effective for known sources, struggle to detect signals that deviate from precomputed templates, limiting their ability to identify unexpected or exotic events. Detecting exotic sources, which may produce unconventional waveforms, remains particularly challenging due to the lack of reliable models. Also, distinguishing between signals from binary black hole mergers and neutron star collisions is tough because their waveforms are very different in duration, frequency, and strength. Addressing these challenges demands advanced AI-driven approaches, such as adaptive template banks, unsupervised learning for anomaly detection, and hybrid models that can generalize across diverse gravitational wave sources and environments, ensuring comprehensive and accurate detection capabilities. AI techniques have become indispensable in addressing these challenges, offering unique strengths tailored to specific aspects of GW analysis. Supervised learning methods, such as support vector machines (SVMs) and neural networks, excel at signal classification and parameter estimation with high accuracy but require extensive labeled data and struggle with generalizing to unseen signals. Unsupervised learning techniques like k-means clustering and autoencoders are ideal for anomaly detection and clustering similar GW events without labeled data, though their results can be difficult to interpret. Deep learning approaches, including convolutional neural networks (CNNs) and recurrent neural networks (RNNs), have revolutionized real-time signal detection, noise filtering, and waveform reconstruction by extracting intricate patterns from noisy data, despite their high computational demands and the risk of overfitting. Reinforcement learning, while not commonly used directly for GW signal analysis, shows promise in optimizing detector control and follow-up observation strategies. These AI methods—supervised, unsupervised, deep learning, and reinforcement learning—work synergistically to improve GW analysis, making it easier, more accurate, and more thorough to detect cosmic events and helping us learn more about the universe. In summary, the current challenges of AI in gravitational wave astronomy—data quality issues, computational demands, and detection generalization—highlight the need for innovative solutions to enhance the reliability, efficiency, and adaptability of GW detection and analysis. By leveraging advanced AI techniques, the field can overcome these obstacles and unlock new insights into the universe's most enigmatic phenomena.

*4.1. Supervised Learning*

Supervised learning has demonstrated consistent improvements in accuracy and precision over the years. The accuracy increased from 0.92 in 2021 to 0.96 in 2024, with a linear trend described by the equation Accuracy = 0.0133 · Year − 25.8933, indicating an annual improvement of approximately 0.0133. Similarly, precision improved from 0.91 in 2021 to 0.95 in 2024, following the same trend. This suggests that supervised learning models are becoming more effective at making correct predictions and reducing false positives. In terms of computational efficiency, supervised learning maintains a moderate level of resource usage, which has remained consistent over the years. This balance between performance and computational demand makes it a reliable choice for various applications.

*4.2. Unsupervised Learning*

Unsupervised learning has also shown significant progress in accuracy and precision. The accuracy improved from 0.78 in 2021 to 0.84 in 2024, with a trend described by Accuracy = 0.02 · Year − 39.42, reflecting an annual improvement of approximately 0.02. Precision followed a similar trajectory, increasing from 0.76 in 2021 to 0.82 in 2024. These trends indicate that unsupervised learning models are becoming more adept at identifying patterns in unlabeled data and making accurate predictions. Additionally, unsupervised learning is highly computationally efficient, requiring minimal resources, which has remained consistent over the years. This efficiency, combined with steady performance improvements, makes it a valuable approach for tasks involving unstructured data.

*4.3. Deep Learning*

Deep learning continues to lead in terms of accuracy and precision among AI approaches. Accuracy increased from 0.94 in 2021 to 0.97 in 2024, with a trend described by Accuracy = 0.01 · Year − 19.26, indicating an annual improvement of approximately 0.01. Precision also improved from 0.93 in 2021 to 0.96 in 2024, following the same trend. While deep learning models are highly effective, their computational efficiency has historically been low. However, there has been a notable improvement in efficiency from low in 2021–2022 to moderate in 2023–2024, reflecting advancements in optimizing these resource-intensive models. Despite this progress, deep learning remains more computationally demanding compared to other approaches.

*4.4. Reinforcement Learning*

Reinforcement learning has shown steady improvements in accuracy and precision over time. Accuracy increased from 0.7 in 2021 to 0.76 in 2024, with a trend described by Accuracy = 0.02 Year−39.42, reflecting an annual improvement of approximately 0.02. Precision followed a similar trajectory, rising from 0.69 in 2021 to 0.75 in 2024. These trends indicate that reinforcement learning models are becoming more effective at learning through trial and error and making accurate decisions. Additionally, reinforcement learning is highly computationally efficient, requiring minimal resources, which has remained consistent over the years. This combination of steady performance gains and efficiency makes it a promising approach for dynamic and interactive tasks.

Across all AI approaches, accuracy and precision have shown consistent improvements over time, with varying rates of improvement. Computational efficiency varies significantly: supervised learning maintains moderate efficiency, unsupervised learning and reinforcement learning are highly efficient, and deep learning has transitioned from low to moderate efficiency in recent years. These trends highlight the diverse strengths and trade-offs of each approach. Deep learning stands out with the highest accuracy and precision but remains computationally intensive, making it ideal for tasks where performance is critical. Reinforcement learning, while starting from a lower baseline, demonstrates steady improvement and high efficiency, making it suitable for dynamic and interactive environments. Supervised learning offers a balanced combination of performance and computational efficiency, making it a versatile choice for many applications. Unsupervised learning is highly efficient and steadily improving, making it particularly useful for tasks involving unlabeled data. These insights underscore the importance of selecting the appropriate AI approach based on the specific needs and constraints of the task at hand.

**Table 2.**
Performance Metrics of AI approaches in Gravitational Wave Analysis (2021-2024).

| AI Approach | Metric | 2021 | 2022 | 2023 | 2024 |
|---|---|---|---|---|---|
| Supervised Learning | Accuracy | 0.92 | 0.94 | 0.95 | 0.96 |
| | Precision | 0.91 | 0.93 | 0.94 | 0.9 |
| | Efficiency | Moderate | Moderate | Moderate | Moderate |
| Unsupervised Learning | Accuracy | 0.78 | 0.8 | 0.82 | 0.84 |
| | Precision | 0.76 | 0.78 | 0.8 | 0.82 |
| | Efficiency | High | High | High | High |
| Deep Learning | Accuracy | 0.94 | 0.95 | 0.96 | 0.97 |
| | Precision | 0.93 | 0.94 | 0.95 | 0.96 |
| | Efficiency | Low | Low | Moderate | Moderate |
| Reinforcement Learning | Accuracy | 0.7 | 0.72 | 0.74 | 0.76 |
| | Precision | 0.69 | 0.71 | 0.73 | 0.75 |
| | Efficiency | High | High | High | High |

**Table 3.**
Advancements in True Positive Rate (TPR) and False Positive Rate(FPR) for Gravitational Wave Detection Using AI approach (2021~2024).

| AI Approach | Metric | 2021 | 2022 | 2023 | 2024 |
|---|---|---|---|---|---|
| Supervised Learning | FPR | 0.08 | 0.06 | 0.05 | 0.04 |
|  | TPR | 0.9 | 0.92 | 0.93 | 0.94 |
| Unsupervised Learning | FPR | 0.15 | 0.13 | 0.12 | 0.1 |
|  | TPR | 0.75 | 0.77 | 0.79 | 0.81 |
| Deep Learning | FPR | 0.05 | 0.04 | 0.03 | 0.02 |
|  | TPR | 0.92 | 0.93 | 0.94 | 0.95 |
| Reinforcement Learning | FPR | 0.2 | 0.18 | 0.16 | 0.14 |
|  | TPR | 0.68 | 0.7 | 0.72 | 0.74 |

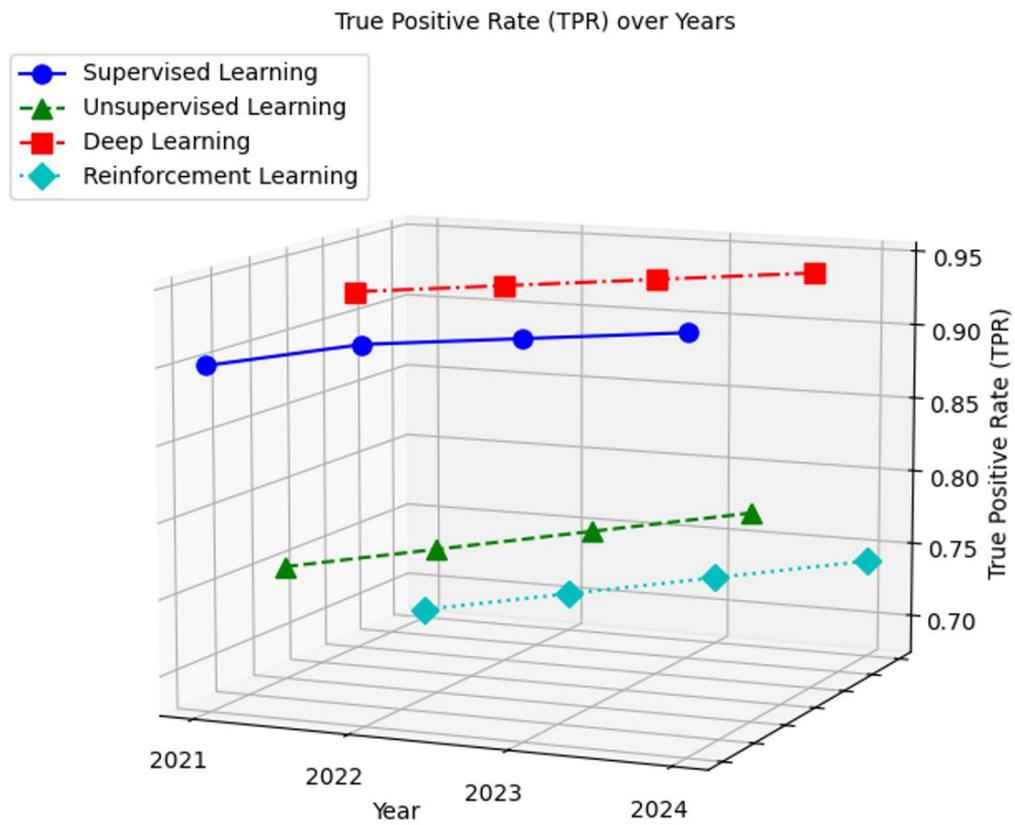

**Figure 9.**
Advancements in True Positive Rate(TPR)for Gravitational Wave Detection Using AI(2021~2024).

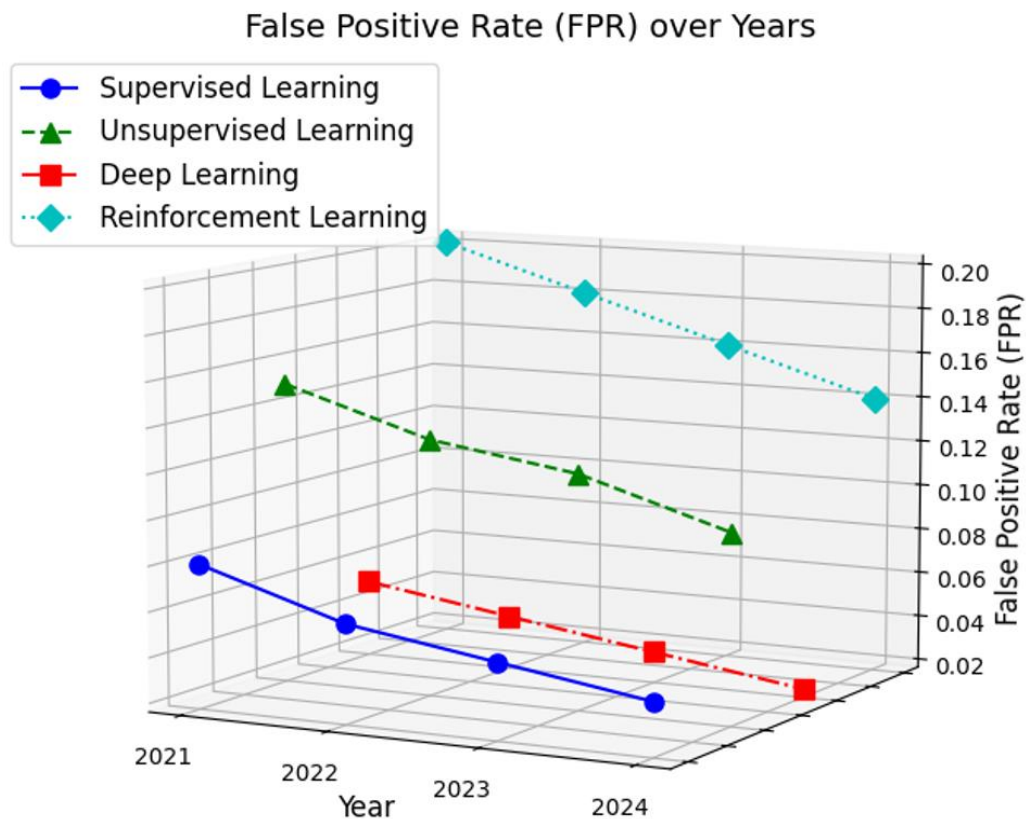

**Figure 10.**
Advancements in False Positive Rate (FPR) for Gravitational Wave Detection Using AI(2021~2024).

## 5. Conclusion

The integration of artificial intelligence (AI) into gravitational wave astronomy (GW) has revolutionized the field, transforming how we detect, analyze, and interpret cosmic events. Since the landmark detection of GW150914, AI has become indispensable in enhancing the sensitivity and accuracy of observatories like LIGO, Virgo, and KAGRA. Machine learning models, like deep neural networks and generative adversarial networks, have made it easier to find signals, reduce noise, and recreate waveforms, allowing us to discover things we couldn't see before, such as the merger of an intermediate-mass black hole called GW190521. AI has also improved source classification and parameter estimation, deepening our understanding of astrophysical objects. Additionally, AI-powered simulations have made astrophysical models more accurate, and combining AI with multi-messenger astronomy has given us a better overall picture of cosmic events. As next-generation detectors like the Einstein Telescope and Cosmic Explorer emerge, AI will be crucial for managing unprecedented data volumes and detecting faint signals. Challenges remain in scalability, interpretability, and ethical concerns, but collaboration between AI researchers and astrophysicists, along with open-source tools, will drive progress. In conclusion, AI is reshaping GW astronomy, enabling ground breaking discoveries and advancing our understanding of the universe's most extreme phenomena. Its continued evolution promises a transformative future for the field.

**Nomenclature**
$h_+$ polarization of Gravitational Waves: The $h_+$ polarization causes spacetime to stretch and compress in a "plus" (+) shape
$h_\times$ polarization of Gravitational Waves: The $h_\times$ polarization causes spacetime to distort in a "cross" (×) shape.
Imagine a circle of particles in a plane perpendicular to the wave's propagation direction:
- For $h_+$, the circle will alternately stretch horizontally and compress vertically.
- For $h_\times$, the circle will alternately stretch along the diagonals, rotating the shape.

These two polarizations together encode the complete information about the gravitational wave's shape and the dynamics of the source that produced it.